\documentclass[osajnl,twocolumn,showpacs,superscriptaddress,10pt]{revtex4-1} 
\usepackage{amsmath,amssymb,graphicx,xcolor}
\usepackage[printonlyused,nohyperlinks,nolist]{acronym}
\usepackage{hyperref}

\newcommand{\be}{\begin{equation}}
\newcommand{\ee}{\end{equation}}

\newcommand{\nvrh}{\; \mathrm{nV}/\sqrt{\mathrm{Hz}}}

\newcommand{\wrh}{\; \mathrm{W}/\sqrt{\mathrm{Hz}}}

\begin{document}
\title{The performance and limitations of \ac{FPGA}-based digital servos for atomic, molecular, and optical physics experiments} 

\author{Shi Jing Yu}
\affiliation{Department of Physics \& Astronomy, University of British Columbia, Vancouver, BC V6T 1Z1, Canada}

\author{Emma Fajeau}
\affiliation{Department of Physics \& Astronomy, University of British Columbia, Vancouver, BC V6T 1Z1, Canada}

\author{Lin Qiao Liu}
\affiliation{Department of Physics \& Astronomy, University of British Columbia, Vancouver, BC V6T 1Z1, Canada}

\author{David J.~Jones}
\email[]{djjones@physics.ubc.ca}
\affiliation{Department of Physics \& Astronomy, University of British Columbia, Vancouver, BC V6T 1Z1, Canada}
\affiliation{Quantum Matter Institute, University of British Columbia, Vancouver, BC V6T 1Z1, Canada}

\author{Kirk W.~Madison}
\email[]{madison@physics.ubc.ca}
\affiliation{Department of Physics \& Astronomy, University of British Columbia, Vancouver, BC V6T 1Z1, Canada}

\date{\today}

\begin{abstract}
In this work we address the advantages, limitations, and technical subtleties of employing \ac{FPGA}-based digital servos for high-bandwidth feedback control of lasers in atomic, molecular, and optical (AMO) physics experiments.  Specifically, we provide the results of benchmark performance tests in experimental setups including noise, bandwidth, and dynamic range for two digital servos built with low and mid-range priced \ac{FPGA} development platforms.  The digital servo results are compared to results obtained from a commercially available state-of-the-art analog servo using the same plant for control (intensity stabilization). 
The digital servos have feedback bandwidths of 2.5 MHz, limited by the total signal latency, and we demonstrate improvements beyond the transfer function offered by the analog servo including a three pole filter and a two pole filter with phase compensation to suppress resonances. We also discuss limitations of our \ac{FPGA}-servo implementation and general considerations when designing and using digital servos.
\end{abstract}


\maketitle

\begin{acronym}[ANOVA]
\acro{AOM}{acoustic optical modulator}
\acro{FPGA}{field programmable gate array}
\acro{PI}{proportional-integral}
\acro{PID}{proportional-integral-derivative}
\acro{PII}{proportional-double-integral}
\acro{PIID}{proportional-double-integral-derivative}
\acro{PI3}[PI\textsuperscript{3}]{proportional-integral-cubed}
\acro{IIR}{infinite impulse response}
\acro{ADC}{analog-to-digital converter}
\acro{DAC}{digital-to-analog converter}
\acro{LSB}{least-significant-bit}
\acro{RC}{resistor-capacitor}
\acro{VO}{voltage offset}
\acro{VGA}{variable gain amplifier}
\acro{PLL}{phase-locked loop}
\acro{SNR}{signal-to-noise ratio}
\acro{IC}{integrated circuit}
\acro{MCU}{micro-controller unit}
\acro{LE}{logic element}
\acro{RAM}{random-access memory}
\acro{ROM}{read-only memory}
\acro{FP}{fixed-point}
\acro{DFI}{direct form I}
\acro{FFT}{fast Fourier transform}
\acro{DSP}{digital signal processor}
\acro{RF}{radio frequency}
\acro{NEP}{noise-equivalent power}
\acro{EOM}{electro-optic modulator}
\acro{ENOB}{effective number of bit}
\acro{P}{proportional}
\acro{GBP}{gain-bandwidth product}
\acro{RIN}{relative intensity noise}
\acro{DF}{direct-form}
\end{acronym}

\section{Introduction}

Active stabilization of physical systems is common in experimental science and ubiquitous in atomic, molecular, and optical (AMO) experiments.  Consequently, feedback-based (i.e.~closed loop) controllers are a fundamental instrumentation element in every AMO physics laboratory.  Typically, stabilization is achieved using proportional-integral-derivative (PID) feedback controllers realized with analog circuitry.  However, significant progress in the speed and power of digital processing units has led to the possibility of replacing analog servos for feedback control with digital controllers.  The advantages of a digital controller over an analog device are many and largely stem from the ability to easily and quickly reconfigure the controller or add complexity to it with firmware or software changes.  The characteristics of a digital controller can be changed dynamically, and therefore adaptive controllers can be easily implemented.  In addition, the complexity of a digital controller is not limited in a practical sense by fabrication time.  This adaptability and potential for high complexity digital controllers enables the realization of features including out-of-lock detection, auto-locking \cite{Leibrandt2015,jorgensen2016simple}, multiple-input and multiple-output transfer functions \cite{Leibrandt2015,sparkes2011scalable,xiao2011digital,schwettmann2011field}, and self-analyzing features \cite{Leibrandt2015,sparkes2011scalable} that would be very difficult and time-consuming to realize with analog filters. 

Previous work on the implementation of digital controllers falls broadly under two categories - the use of computers or microcontrollers for low speed control applications and the use of digital signal processors implemented with field-programmable gate arrays (FPGA) to achieve high bandwidth feedback control. In particular, microcontroller-based feedback systems are typically limited to feedback bandwidths below about 100 kHz \cite{Allard2004,Xu2011,Huang2014}, whereas \ac{FPGA}-based servos have been demonstrated with feedback bandwidths as high as a few MHz \cite{Stockton:02,Jacky2008,Leibrandt2015}. Among the \ac{FPGA}-based servo designs, a few specialized all-digital or partially digital servo designs that forgo the use of an \ac{ADC}/\ac{DAC} have demonstrated bandwidths above 5 MHz \cite{xu2012digital,xiao2011digital}. In spite of considerable development in \ac{FPGA}-based servos, what appears to be missing from the work published thus far is a detailed discussion of the limitations of digital controllers in comparison with their analog counterparts and a side-by-side, quantitative comparison of a digital servo with an analog controller tuned to implement the same transfer function.

Ref.~\cite{Leibrandt2015} presents an \ac{FPGA}-based digital servo with a control bandwidth above 1~MHz capable of multiple-input, multiple-output control optimized for feedback control of lasers in AMO experiments.  The authors also generously share their hardware and firmware design as open source.  They demonstrate auto-locking and transfer function measurement features of their servo in two locking scenarios (frequency control of a laser and length control of a cavity). They note that limitations in servo performance can arise from finite precision (i.e.~fixed point) math, truncation of the filter coefficients and data, and potential register overflow due to high gain.  They also provide a figure comparing the ideal continuous time transfer function and the discrete time, integer math transfer function realized by the \ac{IIR} filters built into their digital servo.  Based on the similarity of the ideal and computed transfer functions, they state that their choice of a 35 bit $\times$ 35 bit signed integer multiplication is sufficient.

The goal of this work is to examine some of the practical limitations and subtleties of realizing a high speed digital servo for AMO experiments and to provide a direct comparison of the lock performance of two different \ac{FPGA}-based servos with a state-of-the-art commercially available analog servo prepared to realize the same transfer function.  Two questions that we  address are what level of \ac{FPGA} is actually needed for a well functioning servo, and what are the noise and bandwidth limiting elements in such a system?  In addition, we discuss several subtle hardware and software issues that we found important to consider in the design and implementation to achieve a high level of performance.  In particular, we discuss the relationship between the coefficient of resolution and distortions in the implemented transfer function and the impact of \ac{LSB} handling in fixed point computations on over-sampling.
We also discuss how we achieved noise suppression in a closed loop control scenario to a level below the analog noise floor of the \ac{ADC}. 

This paper is organized as follows. Section~\ref{sec:hardware} discusses aspects of our hardware design including component selection, latency characterization, and the role of oversampling in the signal processing chain.  Section~\ref{sec:firmware} discusses firmware and software implementation issues including an optimized \ac{IIR} filter design, the role of computational precision, our use of an embedded micro-processor, and our code description.  Section~\ref{sec:baseline} compares the baseline performance of two digital servos (based on the Altera DE2 and DE3 development boards made by Terasic) with a commercial analog controller (Vescent Photonics D2-125 Laser Servo).  We discuss the limitations of the servo bandwidth and noise floor and possible improvements to our design.  Section~\ref{sec:benchmark} presents a side-by-side performance comparison of the digital and analog controllers in a high-bandwidth control application, namely laser intensity stabilization using an acousto-optic modulator.  We also demonstrate two improvements in the \ac{FPGA} servo with the use of \ac{PI3} and \ac{PII} with phase compensation transfer functions.

\section{Design of the \ac{FPGA}-based Servo}

\begin{figure}[h!]

\resizebox{0.48\textwidth}{!}{
  \includegraphics{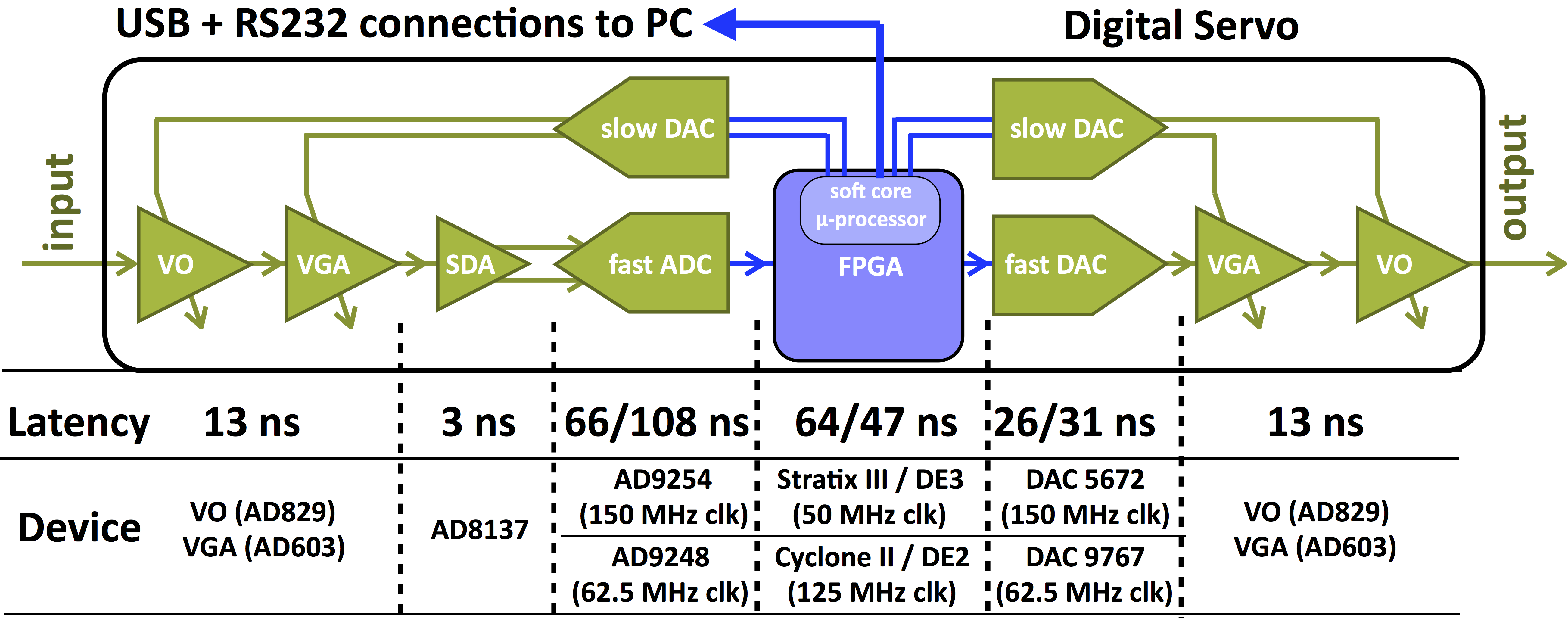}}
\caption{Layout of the \ac{FPGA}-based digital servo. Each servo implements two channels (the second channel is not shown). The analog region of the mixed-signal system is composed of VO (variable offset) circuits, VGAs (variable gain amplifiers) and an SDA (single-to-differential amplifier). Both the input and output VOs and VGAs are controlled by the soft core microprocessor within the \ac{FPGA}. The \ac{FPGA} itself is controlled via RS-232 with a MATLAB GUI running on a PC. The latency of each stage, including the two different \ac{FPGA} platforms denoted DE2 and DE3, is indicated below the schematic. The circuit designs for the VO, VGA, and SDA can be found in Ref.~\cite{yu2017}. 
}
\label{fig:servo-block-diagram}
\end{figure}

A block diagram of our \ac{FPGA}-based servo is shown in Fig.~\ref{fig:servo-block-diagram} \cite{OpenSource}.  To control the servo, an external PC runs a MATLAB script that computes and uploads servo parameters including infinite impulse response (IIR) filter coefficients and servo gain and offset settings for the analog signal conditioning stages.  In the following two sections, we discuss the specific hardware configuration and \ac{FPGA} firmware design.

\subsection{Hardware Configuration}
\label{sec:hardware}

As illustrated in Fig.~\ref{fig:servo-block-diagram}, each \ac{FPGA} servo is composed of two \ac{VO} stages, two \ac{VGA} stages, an \ac{ADC} (including an \ac{ADC} driver), a \ac{DAC} and an \ac{FPGA}. The \ac{VO} stages and the \ac{VGA} stages are components of the analog front- and back-end of the \ac{FPGA} servo and are responsible for tailoring the input and the output of the servo signal to the operating range of the \ac{ADC} and the \ac{DAC}.  Once the signal is converted into the digital domain by the \ac{ADC}, the \ac{FPGA} is responsible for implementing the transfer function, elaborated in Section \ref{sec:firmware}.  Our design uses commercially available \ac{FPGA} development boards, the DE2 and DE3 by Terasic, and fast data acquisitions cards that are compatible with these \ac{FPGA} platforms (the Terasic AD/DA and ADA cards). In our first prototype, we implemented the analog front- and back-end circuits using the AD829 (for the \ac{VO}) and the AD603 (for the \ac{VGA}) on separate boards, and we modified the data acquisition cards to allow DC coupling.  Although some of the data presented here is taken with this first prototype system, we have since developed a single daughter-card that implements all the necessary elements and that is compatible with the DE2, DE3, and the more recent \ac{FPGA} platforms made by Terasic \cite{Terasic}.

As discussed above, we benchmark the performance of servos built with two different \ac{FPGA}s, in part to explore the requirements for developing an \ac{FPGA} servo design.  The DE3's Stratix III \ac{FPGA} is much more powerful than the DE2's Cyclone II \ac{FPGA} in terms of computational resources, with 384 compared to 35 $18\times18$-bit multipliers.   However, for the purposes of the digital servo, this difference between the \ac{FPGA}s is not significant since the most complex filter we implemented in this work (a 3rd order IIR with 32 bit coefficients and 28 bits of fractional resolution) consumed only 18 of the multipliers (for more details on resource allocation, please see Section 3.2.6 of Ref.~\cite{yu2017}).  Also, both \ac{FPGA}s can implement the fast parallel interfaces needed to interface with the \ac{ADC}s and the \ac{DAC}s with clock speeds in the low hundreds of MHz.  Other criteria relevant to our design are the availability of enough \ac{LE}s to implement the soft-core \ac{MCU}, adequate \ac{RAM} for run-time memory of the \ac{MCU}, and onboard \ac{ROM} for the non-volatile storage of the \ac{FPGA} image and \ac{MCU} binary.  Both of the commercial \ac{FPGA} platforms have adequate resources in this regard \cite{yu2017}.

Regarding the \ac{ADC} and the \ac{DAC}, it is essential to clock them at their maximum speed in order to minimize the conversion delay.  The \ac{ADC}/\ac{DAC} combination used for the DE3 servo is the AD9254 and the \ac{DAC}5672, whereas the combination used for the DE2 servo is the AD9248 and the \ac{DAC}9767. Their maximum clock frequencies and pipeline delays along with those for a few other \ac{ADC}/\ac{DAC} combinations suitable for servo designs can be found in Table 2.2 of Ref.~\cite{yu2017}.

We note that these digital components have maximum clock frequencies that are different from the FPGA, and because of the available clock sources and the \ac{PLL} division/multiplication configurations on the development platforms, we clock each component at a multiple of the same base frequency.  On the DE2 platform, the optimal clock speed of 65 MHz cannot be derived from the DE2's base clock source of 50 MHz, and thus the \ac{ADC} (AD9248) is clocked at the slightly lower speed of 62.5 MHz. 

The \ac{VO} and \ac{VGA} stages in the analog front-end circuity are necessary to match the input signal to the fixed voltage range of the \ac{ADC} and thus to fully utilize its dynamic range.  Sampling a signal that extends over only a portion of this range leads to a lower \ac{SNR} than can be achieved with proper signal conditioning.  These same stages are also used in the back-end to provide range control for the output signal after the \ac{DAC} stage.  The \ac{VO} uses the AD829 in an adder configuration with current compensation, and the \ac{VGA} uses the AD603 configured in low-gain mode.  We verified the bandwidth of these circuits to be larger than 70 MHz and to be constant over the \ac{VGA} tunable gain of 40 dB.  These signal conditioning stages thus introduce a small additional delay of approximately 13~ns.

A notable source of noise in our \ac{FPGA} servo originates from the output of the AD603, estimated to be in the range 50-100$\nvrh$.  This noise and the noise from the \ac{VO} stage originates, in part, from the noise of the slow-DACs (DAC7744 used in the prototype design and DAC8734 used in the custom daughter-card design) that are used to control the \ac{VO}s and the \ac{VGA}s.  The \ac{ADC} and the \ac{DAC} also introduce noise into the signal chain above the $10\nvrh$ level.  However, we find that the detrimental effect of the intrinsic output noise of the AD603 (VGA circuit) can be reduced by increasing the \ac{VGA} gain, while the effect of the noise produced by the slow-\ac{DAC} can be mitigated by scaling the voltage down with a resistive divider at the output of the slow-\ac{DAC}.  Clearly, these two noise sources are not inherit to the design and result from our selection of \ac{IC}s used in this design.

The noise and error introduced by the \ac{ADC}/\ac{DAC} is a result of introducing the digital domain.  Quite remarkably, due to over-sampling effects \cite{stewart1998oversampling}, the noise floor of the 14-bit \ac{ADC} used in the DE2 servo is much lower than its quantization step (1 $V/2^{14} \approx 60$ $\mu V$).  In practice, we find the \ac{ADC} noise floor to be limited by the noise in the reference voltage: 136$\nvrh$ at frequencies above 1 kHz and a -20 dB/decade slope (the pink noise) at frequencies below 1 kHz.  Noise levels in the 100$\nvrh$ range can become an obstacle to obtaining an overall noise floor in the 10$\nvrh$ range.  However, as we show in this work, it is possible for an \ac{FPGA} servo to reach a noise floor lower than the analog noise floor of the \ac{ADC} itself by carefully distributing the gain within the system.  A discussion of this technique can be found in Section \ref{sec:noise_floor}.

\subsection{\ac{FPGA} Firmware}
\label{sec:firmware}

The high-speed signal processing tasks, including the servo logic, are handed in the \ac{FPGA} fabric and the slower tasks (communication, slow ramp generation) are handled in the soft-core \ac{MCU}. Detailed information on the Nios II \ac{MCU} used in our firmware can be be found in the Hardware/Software Programmers Manual published by Altera \cite{intel:software,intel:nios2}. In this section, we discuss the servo logic implementation, and we review the critical features of the \ac{IIR} filter implementation for a closed-loop servo application.

\begin{figure}[h!]
\resizebox{0.48\textwidth}{!}{
  \includegraphics{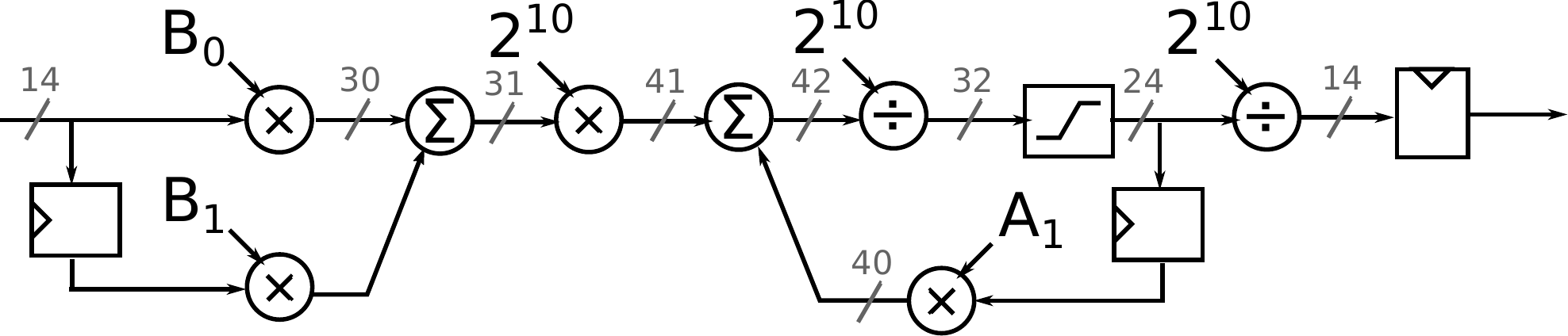}}
\caption{A schematic of a first-order \ac{IIR} implementation (specific to the DE2) with fixed point coefficients ($B_0$, $B_1$ and $A_1$) with 10-bit fractional resolution is shown. The \ac{FPGA} implementation is constructed from simple operations including multiplication ($\times$), addition ($\sum$), and division ($\div$) by the power of 2 implemented as a shift operation, saturation logic (\protect\includegraphics[height=0.35cm]{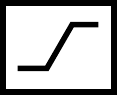}) and clock delays (\protect\includegraphics[height=0.35cm]{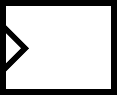}.).
The placement of the multiplication and division by $2^{10}$ operations is critical to correctly implement this \ac{IIR} filter with proper over-sampling.  The bit width of each data path is noted in grey, and we note that the 14 bit width of the input and output signals are limited by the \ac{ADC} and \ac{DAC} conversion stages.}
\label{fig:iir_implementation}
\end{figure}

Figure \ref{fig:iir_implementation} illustrates the implementation of an \ac{IIR} filter with 16-bit coefficients that have a 10 bit fractional resolution.  In this example, the \ac{IIR} filter coefficients are implemented in Q5.10 \ac{FP} number format where 10 specifies the number of fractional bits, 5 is the number of integer bits, and one bit is used to encode the sign of the coefficient.  Here, the \ac{IIR} filter is implemented similarly to the \ac{DFI} of an \ac{IIR} filter (definition of \ac{DFI} can be found in Ref.~\cite{najim2013digital} or textbooks on digital filters) and the \ac{FP} coefficients, annotated in capital letters, are related to the \ac{DFI} definition of the \ac{IIR} coefficients (in lower case letters) as $b_n = B_n/2^{R}$ and $a_n =-A_n/2^{R}$, where $R$ denotes the fractional resolution.

In our design, we reduce computational complexity and increase speed by implementing division and multiplication by powers of 2 as shift operations and by avoiding the need to negate past data in the feedback path (our \ac{IIR} filter implementation differs from the standard \ac{DFI} by the sign of the coefficients in the feedback path).

The bit width of each signal in the \ac{IIR} implementation is annotated in grey in Fig.~\ref{fig:iir_implementation}.  The use of 24 bits for $A_1$, as opposed to the 14 bits available at the \ac{IIR} output, is of particular importance.  Keeping these additional 10-bits in the feedback path of the \ac{IIR} is analogous to keeping 3 more decimal places (in base 10) before the multiplication operation.  This dramatically reduces the rounding error that would have otherwise occurred and is an essential feature to take advantage of over-sampling effects within the IIR filter \cite{stewart1998oversampling}.  It is important to note that this particular placement of the multiplication and division by $2^{10}$ operations is critical to implement the \ac{IIR} filter with proper over-sampling.

The servo implementation on the DE3 platform is similar in layout to the \ac{IIR} filter used in the DE2 servo, shown in Fig.~\ref{fig:iir_implementation}. The main difference is the use of 32-bit coefficients and references to the \ac{DSP} structures that are specific to the Stratix III \ac{FPGA}.  These added bits for the coefficients on the DE3 increase the frequency resolution for the placement of poles and zeros in the transfer function (see Section \ref{sec:tf}).  The references to the \ac{DSP} structure are necessary to route data through the structures built in to the \ac{DSP} that make summation of multiple \ac{DSP} outputs more efficient.  While these \ac{DSP} features are convenient, their use makes the DE3 implementation only compatible with the Stratix III \ac{FPGA}, unlike the platform independent DE2 implementation.

In terms of the computational delay in the servos, we note that the delay through the \ac{IIR} filter is small compared to the total conversion delay. In both the DE2 and the DE3 \ac{IIR} implementation, we are able to trim the computational delay of each third-order \ac{IIR} filter to one clock cycle where the clock frequencies are 50 MHz (DE3) and 125 MHz (DE2). This allows our design to cascade 2 to 3 third-order \ac{IIR} filters while keeping the total signal latency of the servo under 200 ns.

Here, we highlight two important observations regarding the behavior of the \ac{IIR} filters that we implemented.  Firstly, keeping additional \ac{LSB}s in the \ac{IIR} filter's feedback path is necessary to preserve the enhancement obtained by oversampling and to reach performance similar to that of the analog servo. This observation is corroborated by numerical simulations, the details of which can be found in Ref.~\cite{yu2017}.

The second observation is that the saturation logic in our implementation leads to a peculiar problem when a single \ac{IIR} filter is used to realize multiple orders of integration.  In our \ac{IIR} implementation, the saturation logic is a simple clamping logic placed at the \ac{IIR} output (labelled as the \protect\includegraphics[height=0.35cm]{saturation_icon} block in Fig.~\ref{fig:iir_implementation}). This clamp prevents the \ac{IIR} feedback signal from overflowing and the \ac{IIR} output register from wrapping back to zero. This simple clamping logic functions correctly for most use cases of the \ac{IIR} filter. However, for the specific scenario where a single \ac{IIR} filter implements a second- or higher-order integration, the filter fails to correctly hold the output at the rails when the input to the \ac{IIR} filter is a constant non-zero offset.  We believe that the source of the problem is that when the saturation logic activates, the effect is equivalent to adding an additional signal at the location of the saturation logic. While a first-order integration does not interact with this additional signal poorly, a second-order (or higher order) integration of this signal results in an additional ramp-like signal at the output of the \ac{IIR} filter causing the problem described above.  A solution to this problem of realizing higher order integration with this design is to cascade two or more \ac{IIR} filters.  

\section{Baseline Performance of Servos}
\label{sec:baseline}
One of the main objectives of this work is to directly compare the performance of the \ac{FPGA}-based servos with an analog servo. In this section, we present baseline performance levels (of noise floor, bandwidth, and resolution of transfer function) of each servo implementation.  For this comparison, we use a commercially available, high performance analog servo (Vescent Photonics D2-125 Laser Servo).
 
\subsection{Bandwidth}

To characterize the speed of the servos, we measure the frequency of the lowest feedback resonance produced by the servo when used in a closed loop setting without a plant.  This frequency defines what we refer to as the servo bandwidth ($f_{\mathrm{bw}}$) and is related to the total signal latency ($T$) by $f_{\mathrm{bw}} \simeq 1/(2T)$.  This bandwidth measure provides an upper bound on the highest frequency that a servo can correct for and can be inferred from the total signal latency through the servo system.  In practice, the control bandwidth achievable is less than $f_{\mathrm{bw}}$ and is determined by additional latencies or resonances introduced by the plant being controlled.

The latency through an op-amp is determined by its \ac{GBP} and the slew-rate of the op-amp.  With the high \ac{GBP} of presently available op-amps, it is not surprising for an analog servo, like the Vescent D2-125, to have a bandwidth higher than 10 MHz.  Like the analog servo, the \ac{FPGA}-based servo has analog signal conditioning stages, but it also has two additional sources of delay that further reduce the bandwidth, the signal conversion delay in the \ac{ADC} and the \ac{DAC} and the computational delay in the \ac{FPGA}.

\begin{figure}[h!]

\resizebox{0.48\textwidth}{!}{
  \includegraphics{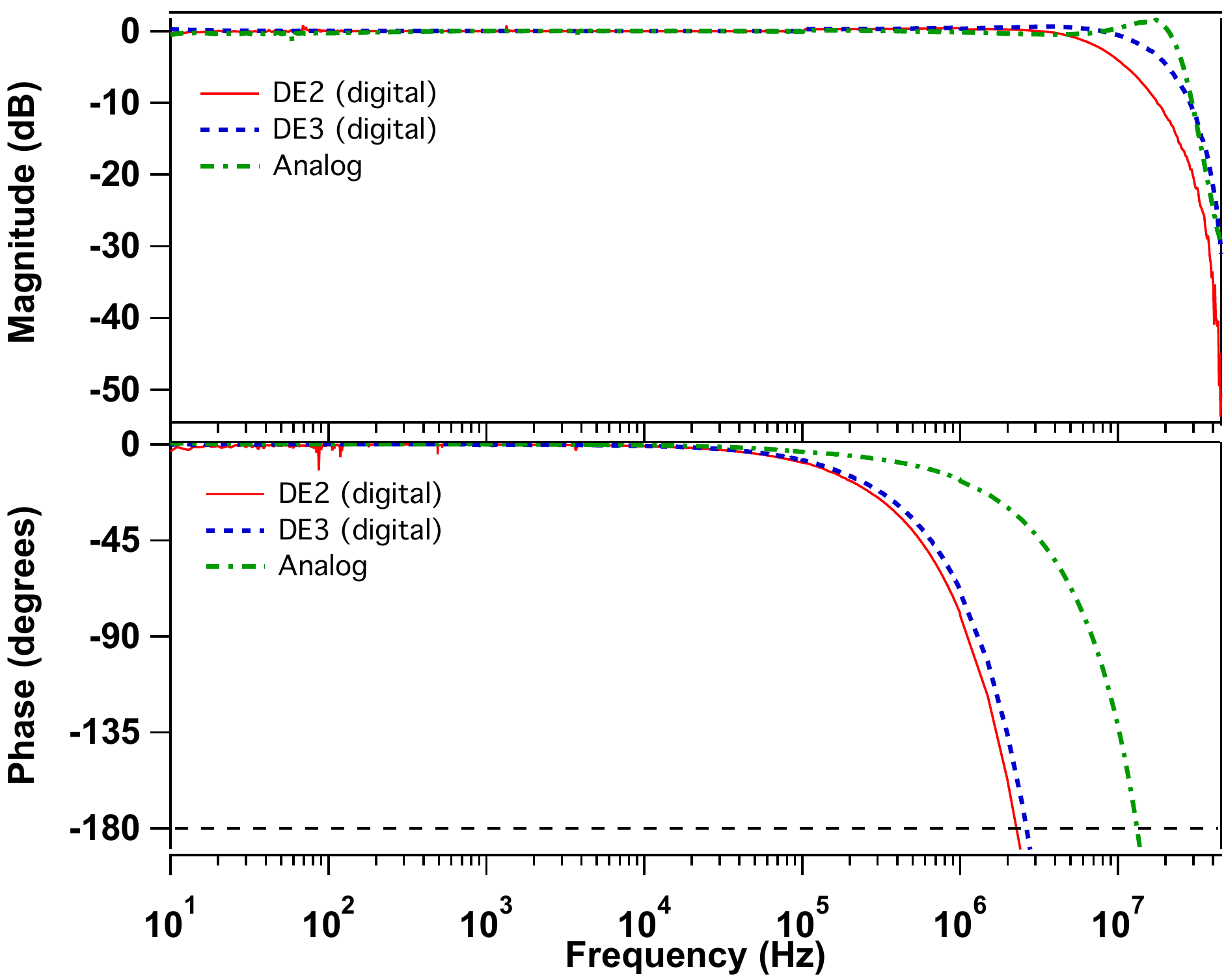}}
\caption{(color online) Transfer function of the servos when configured as proportional controllers and measured with SR780 (up to 100 kHz) and HP8753E (100 kHz to 40 MHz). The only major difference is the bandwidth of the servos as depicted by the intersection between the phase response of each servo and the 180 degree phase shift boundary (the black dashed line).
}
\label{fig:servoBW}
\end{figure}

The bandwidths of the servos are measured with a Stanford Research Systems SR780 Network Signal Analyzer (from DC to 100 kHz) and with a 
Hewlett Packard HP 8753E Network Analyzer (100 kHz to 40 MHz) with each servo configured for unity gain with only proportional feedback. For the digital servos, all \ac{IIR} filter computations are still executed (3 cascaded \ac{IIR} filters all configured with unity gain) to properly quantify the effect of the computational delay.  The measured transfer functions for these \ac{P} controllers are shown in Fig.~\ref{fig:servoBW}.  These measurements verify the specified 10~MHz bandwidth of the Vescent D2-125 and show that the \ac{FPGA} servo has the 2.5~MHz bandwidth expected from the 200~ns signal latency.  This latency is mainly due to the additional conversion delay introduced by the \ac{ADC} and the \ac{DAC} and, to a smaller extent, the computational delay in the \ac{FPGA} as illustrated in Fig.~\ref{fig:servo-block-diagram}.

As illustrated in Fig.~\ref{fig:servo-block-diagram}, the signal latency through the \ac{FPGA} servo is composed of several delay sources. The conversion delay is 92(139) ns and accounts for 60(75)\% of the total delay in the DE3(DE2) \ac{FPGA} servo. The computational delay takes up a smaller fraction as a result of the speed optimization in designing the \ac{IIR} logic, with the cascade of two third order \ac{IIR} filters each costing 64(47)~ns for the DE3(DE2) servo and accounting for 35(22) \% of the total delay. The total delay through the analog signal conditioning stages in this design is only 30 ns (15\% of the total) and is on the order of the total delay through an analog servo with similar op-amp technology.  The delay in signal conversion and, to a lesser extent, the computational delay are the primary culprits in the additional delay in an \ac{FPGA} servo.

Does the additional delay in a digital servo limit its application?  Surely, because of the longer signal delay of the \ac{FPGA} servo, the bandwidth of a control system employing the \ac{FPGA} servo cannot exceed 2.5 MHz.  However, because the bandwidth of a control system also depends on the signal latency through the rest of the system, the difference between the \ac{FPGA} servo and the analog servo may not be substantial when they are used to control a plant that has a considerably longer latency.

\subsection{Noise Floor}
\label{sec:noise_floor}

In this section, we discuss noise floor measurements of each servo.  These measurements indicate the minimum signal level that the servo can correct for. As a benchmark for this comparison, the analog servo is specified to have a noise floor of 10$\nvrh$ by the manufacturer, and this value is confirmed by our measurements.

To reveal the noise floor relevant to the servo performance in a control scenario, we set the transfer function to \ac{PII} with corner frequencies at 200 kHz and configure it in a closed loop by connecting the output directly to the input.  We refer to this as a self-locking configuration. The rationale for measuring the noise floor in this self-locking (closed-loop) configuration rather than configuring the servo in open loop with unity gain is to reveal the input noise floor level of the servos.  The input noise level is much more important than the output noise level since for any control system the noise at the output of the servo loop is suppressed across the servo bandwidth where the servo gain is large.

\begin{figure}[h!]
\resizebox{0.48\textwidth}{!}{
\includegraphics{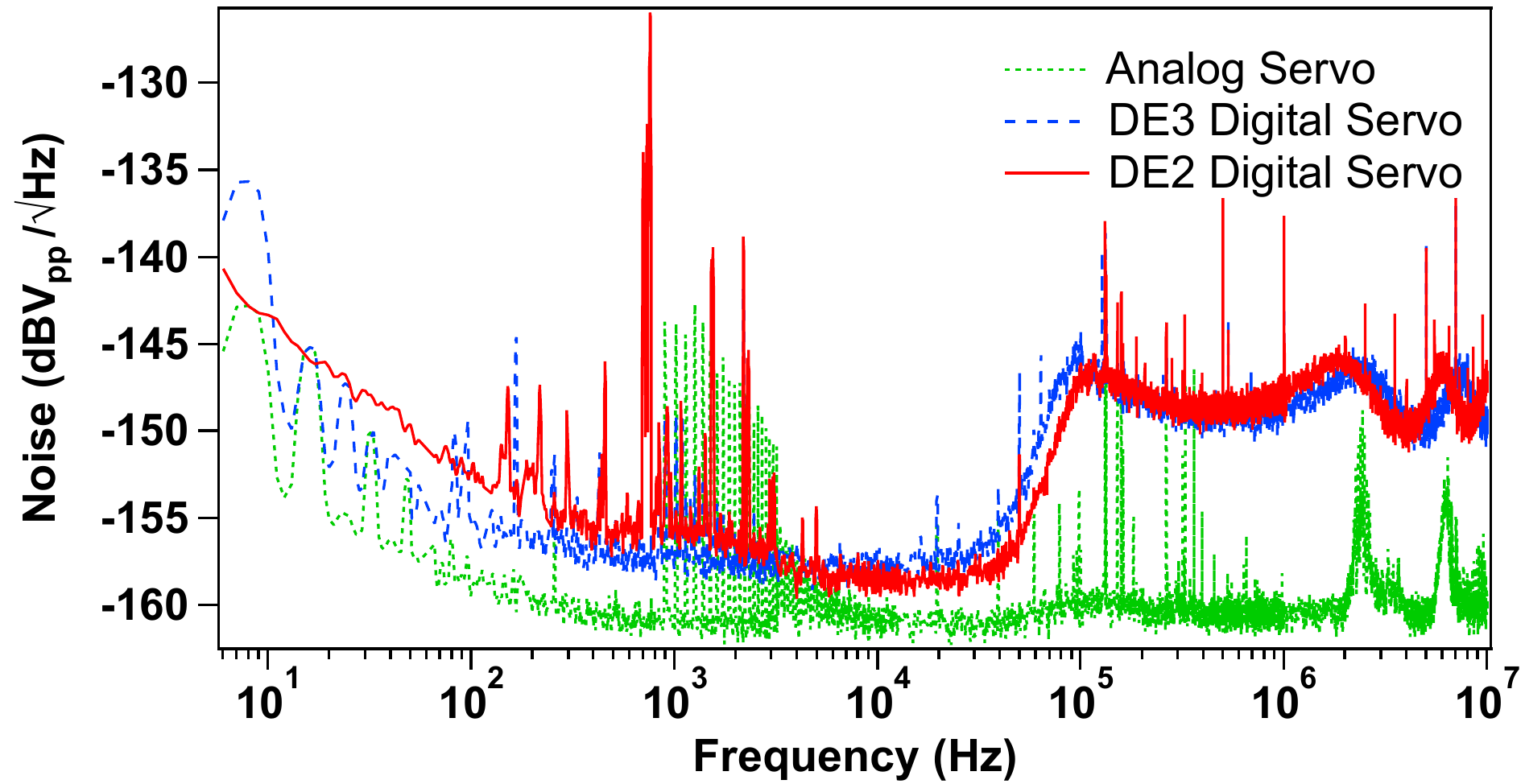}}
\caption{(color online) The noise floor comparison of the servos. The servos are configured as PII controllers in closed-loop with the same total gain at high frequencies.  The gain distribution of the analog servo was set according to the manufacturer's
specifications; however, because the \ac{FPGA} servo has noise sources on the input stage above 10$\nvrh$, the input signal gain is maximized before the \ac{ADC} and then attenuated either in the \ac{FPGA} logic or after the \ac{DAC} accordingly to produce the same total gain as the analog servo required for a direct comparison.  This strategy yielded similar performance to the analog servo at frequencies lower than 30 kHz.
}
\label{fig:self-locking}
\end{figure}

The results of the noise floor comparison are shown in Fig.~\ref{fig:self-locking}.  The difference in the noise levels between the analog servo and the \ac{FPGA} servos for frequencies below 30 kHz is less than 5 dB.  We will refer to this low frequency region (frequencies below 30 kHz) as the suppression band in the rest of the manuscript. It is remarkable that the noise performance for the \ac{FPGA} servos in this self-locking configuration is below the noise floor of the \ac{ADC} itself.  This was achieved by maximizing the input signal gain while keeping the total loop gain of the servo fixed.  For frequencies higher than 30 kHz, the \ac{FPGA} servos have a higher noise level by at most 15 dB.  As discussed in Sec.~\ref{sec:firmware}, the \ac{FPGA} servo noise level (in the range 50 to 100~$\nvrh$) is larger than that of the analog servo ($\sim 10 \; \nvrh$) due to noise introduced by the \ac{VGA}, the \ac{ADC}, and the \ac{DAC}.  This noise arises, in part, from voltage noise on the reference voltage to the \ac{ADC} and voltage noise of the slow-\ac{DAC} used to control the analog front-end settings.

For closed-loop applications, the noise levels at different locations of the loop have a very different impact on the output of the closed-loop system. Noise at the input of the servo is directly written onto the output of the system, while the noise at the output of the servo is suppressed due to the action of the servo itself. To achieve the noise performance in the suppression band shown in Fig.~\ref{fig:self-locking}, we maximized the gain before the \ac{ADC} and decreased the gain after the \ac{DAC} accordingly (with a fixed attenuator) to keep the total loop gain constant and equal to the total loop gain of the analog servo.  We found that decreasing the gain in the \ac{FPGA} logic can also be used instead of reducing the gain after the \ac{DAC}.  This latter method has the advantage of maintaining the full voltage range of the servo output.  This is often needed to correct a disturbance of a fixed amplitude.  However, reducing the gain inside the servo logic must be done with care, because the noise floor achievable by the logic and resolution of the transfer function can be affected in the process (see Ref.~\cite{yu2017} for more details).

In addition to the noise sources in the analog domain, the digital implementation of the servo has an effect on the noise floor. As mentioned previously, correctly handling the \ac{LSB}s in the \ac{IIR} implementation is critical to achieving the best noise floor. In particular, we found that degradation of the signal quality in the digital logic implementation often resulted from incorrect rounding or truncation .

\subsection{Transfer Function}
\label{sec:tf}

In this section, we address the range of transfer functions that a servo can implement. Typically, analog servos are designed with a discrete set of \ac{RC} values to implement a discrete set of corner frequencies between 10 Hz and 1-2 MHz. The \ac{FPGA} servo also has a discrete distribution of poles and zeros due to the underlying \ac{IIR} implementation. In a first order \ac{IIR} filter, the spacing between adjacent poles and zeros is roughly constant and equal to $\Delta f = \alpha f_{\mathrm{clk}}/2^R$, where R is the fractional resolution of the coefficient (its width in bits), $f_{\mathrm{clk}}$ is the clock frequency, and $\alpha$ is the scaling factor which for this case is $\alpha = (2 \pi)^{-1}$ \cite{yu2017_section}. For the implementation with the \ac{IIR} coefficients in Q5.10 format (16-bit signed coefficients with a 10 bit fractional resolution) and an $f_{\mathrm{clk}}$ of 50 MHz, the frequency resolution is 7.7 kHz.  By contrast, for the implementation with the \ac{IIR} coefficients in Q3.28 format (32-bit signed coefficients with a 28 bit fractional resolution), the resolution is 48 mHz. It is possible to improve the pole/zero resolution of the \ac{IIR} filter with 16-bit coefficients by slowing the clock, but we did not pursue this approach due to the additional latency that would result. 

The frequency resolution of a high-order \ac{IIR} filter is more complex, since the frequency resolution of the zeros/poles affect one another \cite{oppenheim1972effects}.  However, it can be shown that a high-order \ac{IIR} filter will, as a result of this correlation, have less precision in placing multiple poles/zeros than placing a single pole/zero. 

\begin{figure}[h!]
\resizebox{0.48\textwidth}{!}{
  \includegraphics{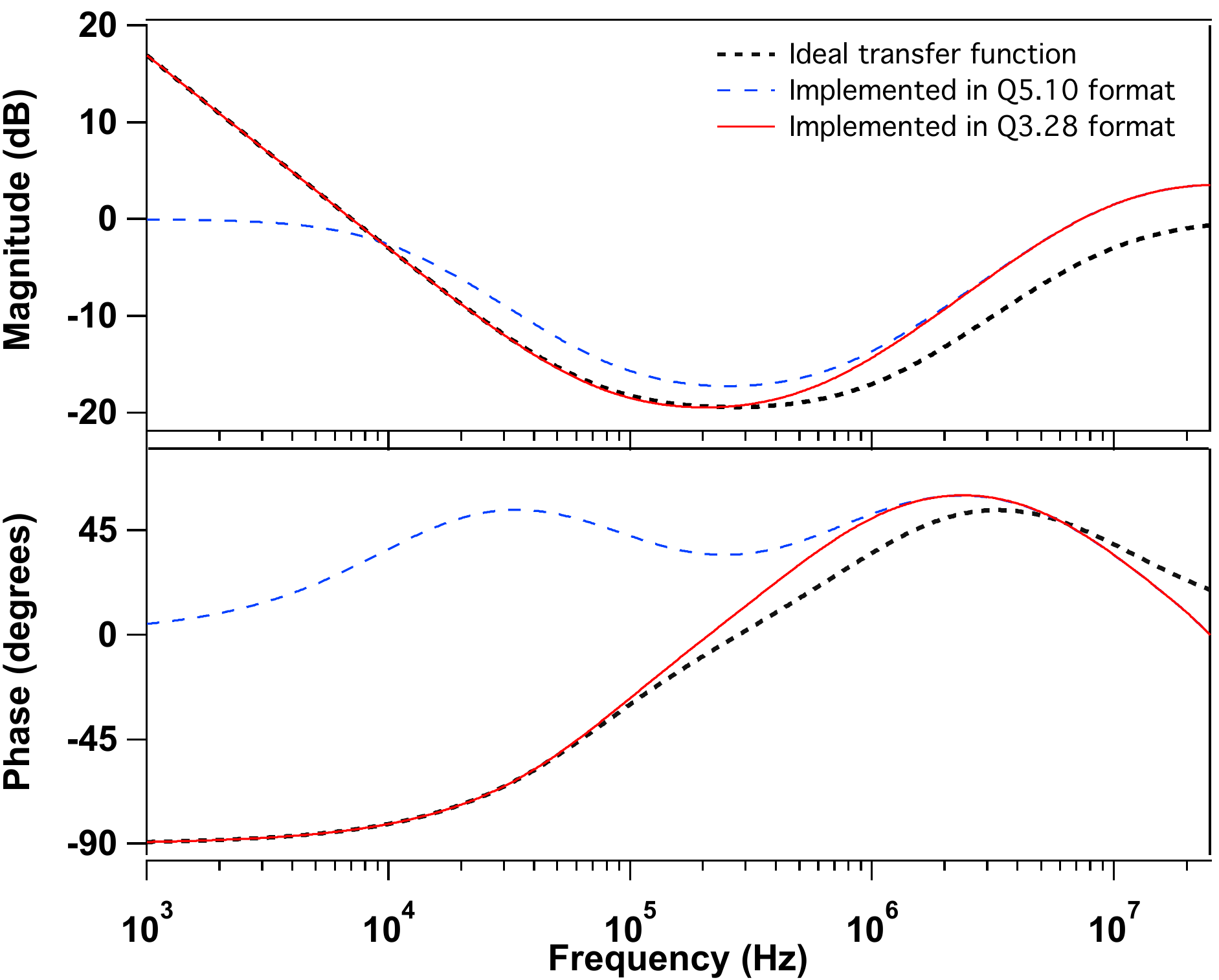}}
\caption{(color online) Numerical simulation of the transfer function that results from an \ac{IIR} filter implementation of a \ac{PID} controller with two different coefficient resolutions: Q5.10 (in blue) and Q3.28 (in red).  In this particular case, the Q5.10 implementation has an unacceptable amount of distortion at low frequencies and is dynamically unstable due to the rounding of the 
IIR coefficients to the nearest hardware realizable value and the resulting movement of the location of the pole/zero from just inside to just outside the stability region in the z-domain.}
\label{fig:freq-resolution}
\end{figure}

Another effect that arises from the finite coefficient resolution is an inevitable distortion of the transfer function.  An example transfer function and its realization by an \ac{IIR} filter with two different coefficient resolutions is shown in Fig.~\ref{fig:freq-resolution}.  Distortions in the magnitude and phase are evident in both cases.  Not only can distortions affect the locking performance of the servo, but they can also lead to dynamical instabilities due to the movement of the pole and zero locations from just inside to just outside the stability region in the z-domain when the IIR coefficients are rounded to the nearest hardware realizable value.  This effect is described in various texts (see for example section 9.4 of Ref.~\cite{mitra2006digital}) and continues to be a topic of study in control research \cite{hilaire2007unifying}.

There are many strategies for reducing the effects of finite coefficient resolution on the accuracy of the implemented transfer function. Cascaded \ac{IIR} filters and \ac{IIR} filters with lattice structures are two examples \cite{szczupak1978digital,mitra2006digital}. For complex transfer functions, including the \ac{PII} with an integrated lag-lead filter discussed in Section \ref{sec:ll}, we use a set of cascaded higher-order \ac{IIR} filters in \ac{DF} with 32-bit coefficients to implement the desired function with high fidelity and good resolution for feature placement. Because the use of high resolution \ac{IIR} filters does not completely eliminate the possibility of distortion, we use a MATLAB simulation to check for discrepancies between the desired transfer function and the one implemented in the \ac{FPGA} servo. 

\section{Benchmark Tests}
\label{sec:benchmark}

In this section, we discuss the use of the \ac{FPGA} servos in an intensity stabilization apparatus (also referred to as a noise-eater) and compare their performance against the commercial analog servo (D2-125) characterized in Sec.~\ref{sec:baseline}.  We also demonstrate the realization of transfer functions with the \ac{FPGA} servo that are more complex than that of the commercial analog servo, and we find significant performance improvement of the noise-eater with the use of \ac{PI3} and \ac{PII} with phase compensation transfer functions.

\subsection{Experimental Setup}

\begin{figure}[h!]
\resizebox{0.48\textwidth}{!}{
  \includegraphics{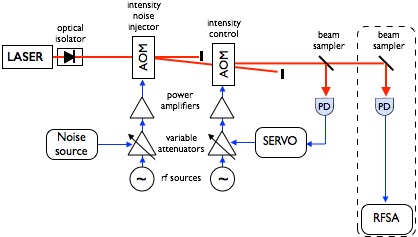}}
\caption{Experimental set-up of the intensity stabilization (noise-eater) system. An arbitrary intensity noise is written on to the laser and then stabilized through a closed-loop formed by the servos used in this study, an \ac{AOM} (acoustic-optical modulator) used as the modulator and an PD (photo-detector) used as a feedback detector. The out-of-loop detection (in dashed box) is used to verify that the intensity noise is truly suppressed and is analyzed by a RFSA (radio frequency spectrum analyzer).}
\label{fig:experimental setup}
\end{figure}

As shown in Fig.~\ref{fig:experimental setup}, the intensity stabilization system is constructed from an \ac{AOM} (IntraAction ATM series) and is controlled by a \ac{RF} source with a variable power realized with a variable attenuator (MACOM MAAVSS0006) nested in the \ac{RF} amplifier chain. The detection of the optical intensity is done with a biased Si photodetector (Thorlabs DET10A) with a \ac{NEP} of $1.2 \times 10^{-13} \wrh$. The active components of the intensity stabilization system including the \ac{AOM}, the variable attenuator, and the photo-detector all have fast rise/fall times ($<$70 ns, 10 ns and 1 ns  respectively). This leads to a relatively high open-loop bandwidth (3 MHz); however, the delay of the action of the \ac{AOM} due to the propagation time of the acoustic wave from the piezo-electric element to and across the laser beam inside the \ac{AOM} adds a signal latency of 400~ns to the closed loop delay of the intensity stabilization system.  This latency is the dominant delay, and therefore the closed-loop performance difference of the system due to the bandwidth difference of the two types of servos is not substantial.

A second \ac{AOM} is used to inject additional intensity noise into the laser light, and this is accomplished by modulating the \ac{RF} power using the voltage controlled attenuator and a voltage noise source produced by an arbitrary waveform generator (Stanford Research Systems DS345).  This set-up allows us to customize the spectrum of the noise and allows a more thorough study of the noise suppression capability of the servos.  In particular, we inject a wide-band ($>10$~MHz), white noise spectrum with a magnitude large enough that the noise suppression achieved by the full system system is far above and therefore not limited by the shot-noise of the detection system.  This is done to insure that the comparisons we are making between servos is representative of their noise suppression capability and not limited by shot-noise.  In addition, while monitoring the output of the photodetector in the servo loop (in-loop detection) provides some evidence of noise suppression, an additional photodetector is used to perform an independent, out-of loop detection of the intensity noise after the noise-eater.  We note that all of the data presented here is from an out-of-loop measurement except for the data on the performance of the \ac{PII} with lag-lead phase compensation.

\subsection{PII}

\begin{figure}[h!]
\resizebox{0.48\textwidth}{!}{
  \includegraphics{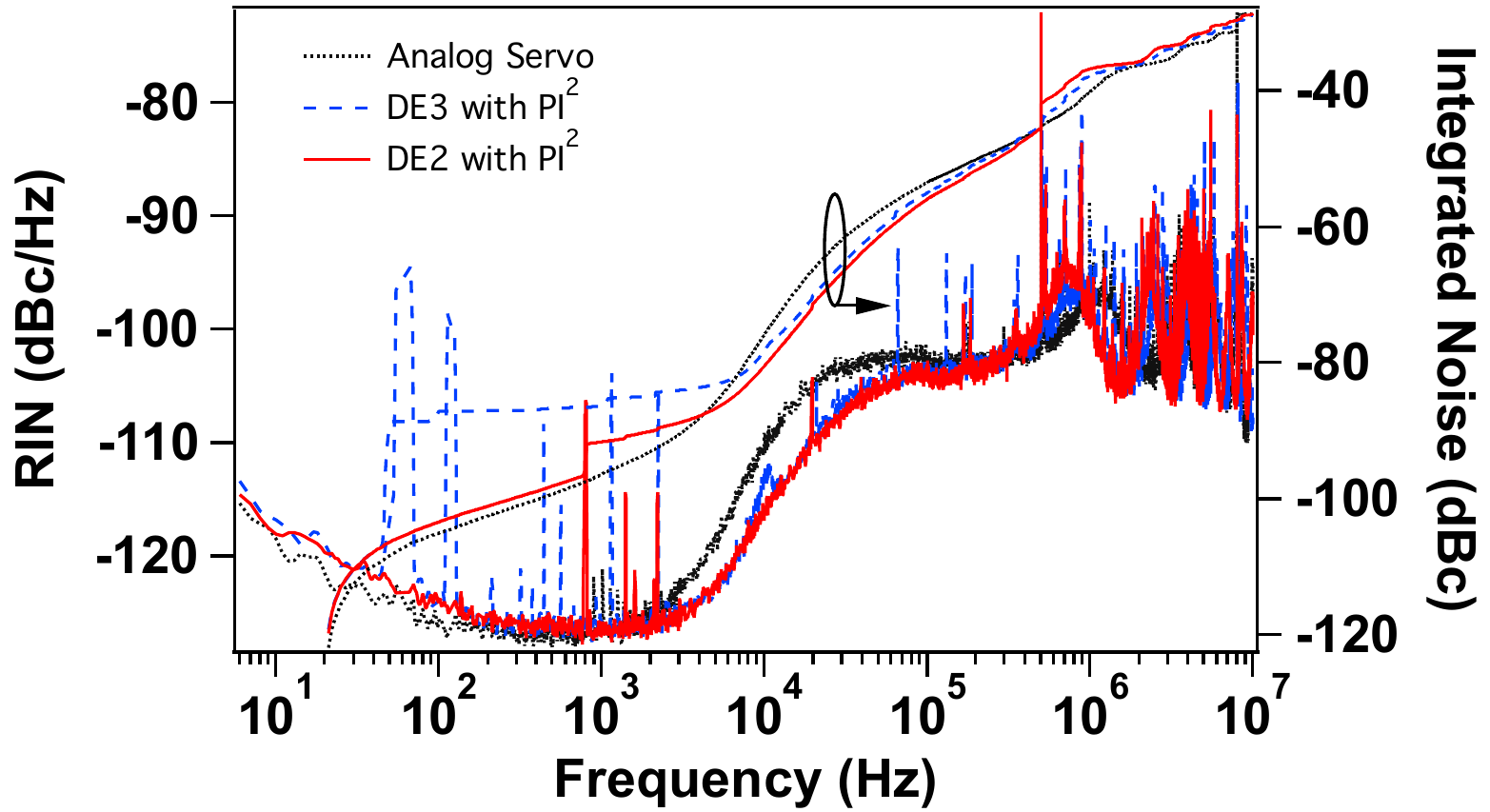}}
\caption{(color online) Comparison of the relative intensity noise spectra (thick lines referenced by the y-axis scale on the left) achieved with the intensity stabilization system using the three different servos controlling the same plant.  The thin lines represent the integrated noise (referenced by the y-axis scale on the right) from 20 Hz up to a given frequency.}
\label{fig:servo_PII}
\end{figure}

The performance of the intensity stabilization system is characterized by the \ac{RIN} of the output light and is shown in Fig.~\ref{fig:servo_PII} with each servo configured as a \ac{PII} controller.  The width of the suppression band is set by the highest corner frequency of the \ac{PII} transfer function. This corner frequency is set to 100 kHz for the analog and 70 kHz for the \ac{FPGA} servos.  Higher values worsen the loop resonances for each servo system.  The analog servo has a small advantage in that it allows a higher corner frequency due to its smaller signal latency.  The second corner frequency is 10~kHz (7 kHz) and is chosen to be a decade below the higher corner frequency for the analog (\ac{FPGA}) servo.  The gain setting for each servo is optimized for each measurement. 

Figure \ref{fig:servo_PII} shows that the \ac{RIN} of the stabilized laser systems are similar both inside and outside the suppression band.  This result is different from the noise floor measurement of the servo in Section~\ref{sec:noise_floor}, where the servos have similar noise floors inside and different noise floors outside the suppression band.  In this system, the noise outside of the suppression band is dominated by the noise injected by the \ac{AOM}.  Because the \ac{PII} controller is incapable of suppressing this out-of-band noise, the noise floor difference between the \ac{FPGA} servo and the analog servo does not cause a visible difference.  In addition, unlike the self locking data, here the loop resonance frequencies are almost the same (700 kHz and 1 MHz).  This is due to the dominant loop delay introduced by the action of the \ac{AOM} (about 400 ns).

\subsection{PI\textsuperscript{3}}

\begin{figure}[h!]
\resizebox{0.48\textwidth}{!}{
  \includegraphics{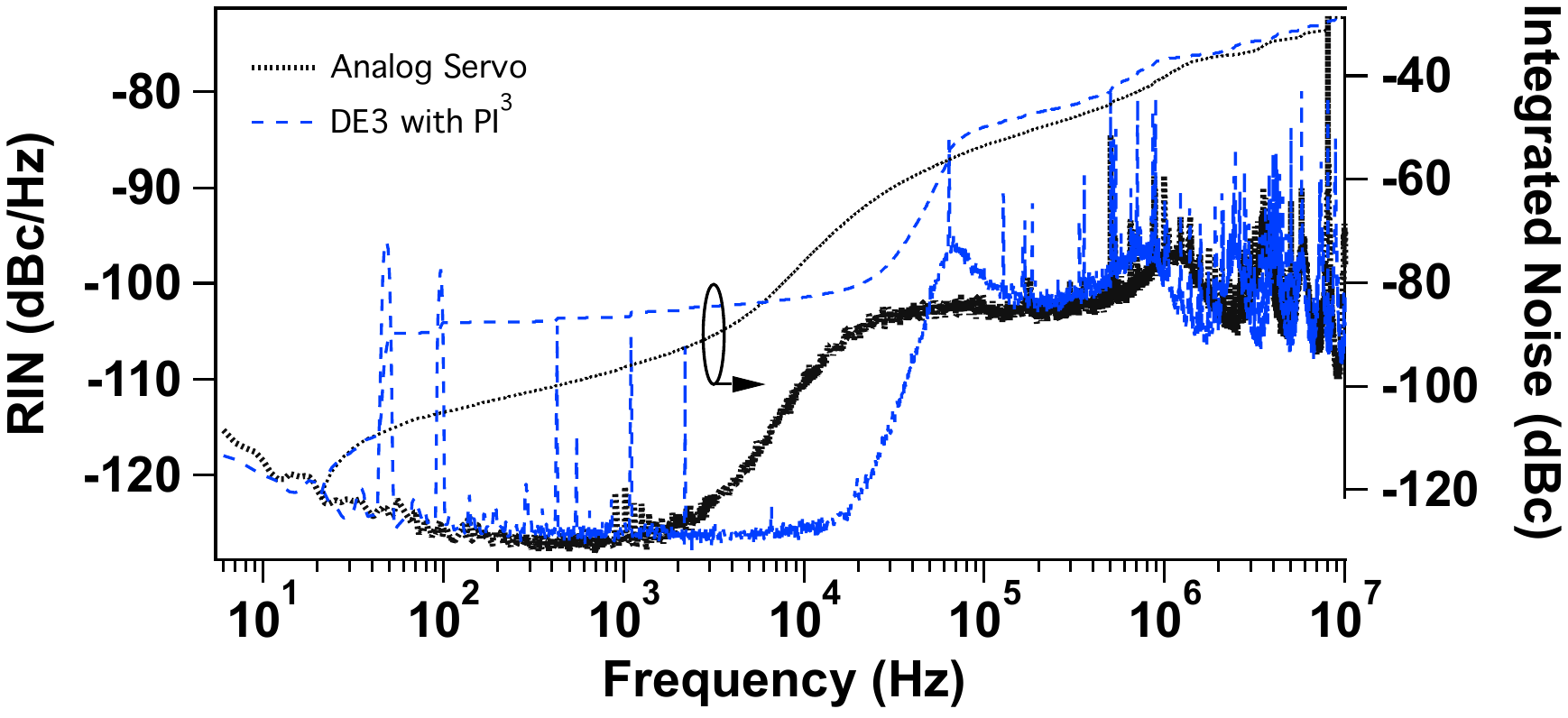}}
\caption{(color online) Relative intensity noise spectra (thick lines referenced by the y-axis scale on the left) achieved with the intensity stabilization system using the analog servo and the \ac{FPGA}-based servo with different transfer functions.  The thin lines represent the integrated noise (referenced by the y-axis scale on the right) from 20 Hz up to a given frequency.  The analog servo realizes a PI$^2$ with the two poles located at 10~kHz and 100~kHz, whereas the digital servo is configured to realize a PI$^3$ with three poles located at 10~kHz, 100~kHz, and 100~kHz.  We note that while this significantly increases the suppression band, this comes at the cost of additional amplitude at 70~kHz.}
\label{fig:servo_PI3}
\end{figure}

As a natural extension of the \ac{PII} servo, a \ac{PI3} filter is implemented for the intensity stabilization system. This is accomplished by cascading three \ac{IIR} filters in the \ac{FPGA} servo, all configured as integrators.  The performance of a \ac{PII} filter realized by the analog servo and a \ac{PI3} filter realized by the \ac{FPGA}-based servo is compared in Fig.~\ref{fig:servo_PI3}.  The \ac{PI3} filter widens the suppression band without increasing the oscillation amplitude at the \ac{FPGA}-based servo loop delay resonance.  This is an example of a filter that cannot be implemented in the analog servo without additional hardware, while it can be implemented in the \ac{FPGA} servo with ease. 

\subsection{PII+LL}
\label{sec:ll}
The lag-lead filter is another example of a filter that cannot be easily implemented in an analog servo. Unlike the \ac{PI3} filter, that can be constructed by chaining together multiple analog servos, the lag-lead filter is rarely implemented in a general-purpose analog servo due to the additional impedance networks that are necessary to make all the poles and zeros tunable.

\begin{figure}[h!]
\resizebox{0.48\textwidth}{!}{
  \includegraphics{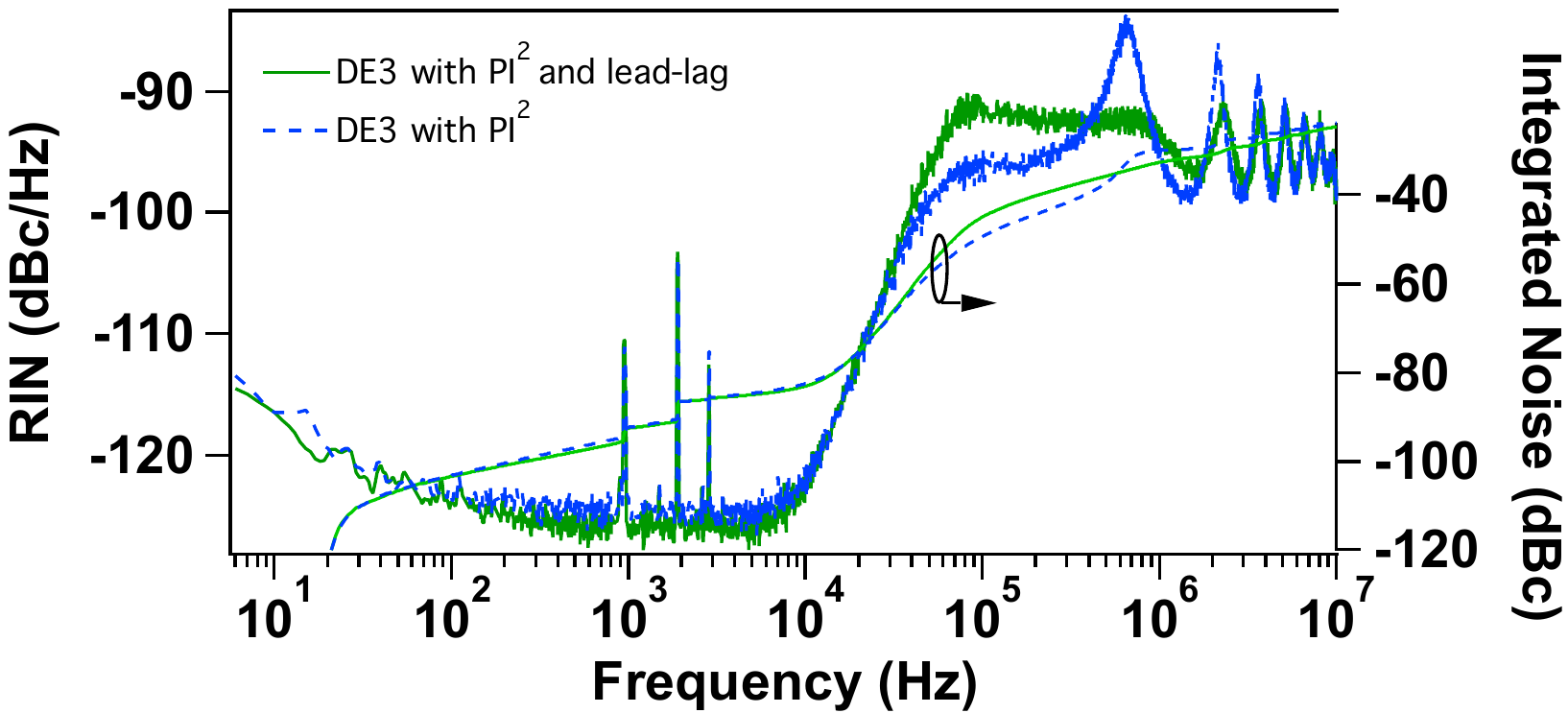}}
\caption{(color online) Relative intensity noise spectra (thick lines referenced by the y-axis scale on the left) achieved with the intensity stabilization system using the \ac{FPGA}-based servo configured with two different transfer functions : a  \ac{PII} and a \ac{PII} with lag-lead.  The thin lines represent the integrated noise (referenced by the y-axis scale on the right) from 20 Hz up to a given frequency.  The lag-lead is a notch filter with a center frequency of 700 kHz.  The lag-lead filter suppresses oscillations at the loop resonance of the servo system.}
\label{fig:servo_PII_LL}
\end{figure}

The lag-lead filter is effectively a notch filter composed of two poles and two zeros. This is not to be confused with the lead-lag filter, although both can be used as phase compensators in different applications. A summary of the two types of phase-compensators can be found in various sections of Chapter 9 of Ref.~\cite{golnaraghi2010automatic}. Here, the lag-lead filter is used to suppress the loop resonance of the \ac{FPGA}-based servo at 700 kHz. The effect of this notch filter on the loop resonance is demonstrated in Fig.~\ref{fig:servo_PII_LL}.  Because the lag-lead filter reduces the gain locally at the loop resonance, the amplitude of the loop oscillation is reduced.

We note that the \ac{PII} + lag-lead transfer function contains a total of 4 poles and 4 zeros and is very susceptible to distortions of the transfer function when the \ac{IIR} coefficients are rounded. This issue is addressed by distributing the low frequency zeros in the \ac{PII} in separate \ac{IIR} filters that are cascaded and by checking with numerical simulations for discrepancies between the intended transfer function and the implemented one.

\section{Conclusions}

In this work, we present two \ac{FPGA} servo designs using the Altera DE2 and DE3 development boards made by Terasic that achieve a total signal latency of 200~ns.
As discussed, the DE3's Stratix III \ac{FPGA} is a factor of ten more powerful (and more expensive) than the DE2's Cyclone II \ac{FPGA} in terms of computational resources; however,
for the purposes of implementing the digital servos explored in this work, the resources on the low-end DE2 \ac{FPGA} were completely sufficient.  We provide benchmark performance comparisons of noise, bandwidth, and dynamic range of these \ac{FPGA} servos with a commercial analog servo.  We also provide a direct comparison of the lock performance of the \ac{FPGA}-based servos with the analog servo using the same plant for intensity stabilization of a laser beam.  For this control application, we also demonstrate the realization of more complex transfer functions including a \ac{PI3} and a PII with lag-lead phase compensation to suppress the loop resonance of the intensity noise system.  We also discuss the role of gain distribution in the servo and how maximizing the input gain while keeping the total loop gain constant is a viable technique for achieving noise suppression to a level below the analog noise floor of the \ac{ADC}.

We also discuss several subtle hardware and software issues that we found important to consider in the design and implementation to achieve a high level of performance with the \ac{FPGA} servo.  In particular, we discuss the primary sources of noise and latency in our design and how these could be improved, and we discuss several subtle aspects of floating-point arithmetic and how they affect the behavior of \ac{IIR} filters.  We discuss transfer function distortion and frequency quantization of poles and zeros arising from \ac{FP} representation of the \ac{IIR} filter coefficients as well as the the role of \ac{LSB} handling in \ac{FP} computations and its impact on over-sampling as it pertains to signal processing.

Some observations made in this work that we wish to reiterate here are as follows.  First, the \ac{ADC} is the largest contributor of noise in our servo design. Its noise floor in the low 100$\nvrh$ range is quite large compared to that of standard op-amps used in analog servos (typically in the low 10$\nvrh$ range).  Due to over-sampling, an effect that requires a small amount of white noise perturbing the input of the \ac{ADC}, this noise floor is well below (here, on the order of 500 times smaller than) the quantization step of the \ac{ADC}.  While the noise floor of \ac{ADC} devices does decrease as their resolution increases, this comes at the cost of additional delay and thus a loss of bandwidth of the servo.  As a result, we find that the optimal performance is achieved by maximizing the input gain while keeping the total loop gain constant.  Second, we find that because over-sampling also plays an important role in the \ac{FP} math of the \ac{IIR} filter itself, the servo logic itself can lead to degradation of the signal if handled incorrectly.  Third, we wish to reiterate that the complexity of the transfer function that a digital servo can realize is limited by the fractional resolution of its \ac{IIR} coefficients.  When the coefficients do not have sufficient resolution, a significant discrepancy between the implemented transfer function and the desired transfer function may result, including the exact placement of poles and zeros.

Finally, we note that the bandwidth limitations of our \ac{FPGA} servo are primarily limited by the conversion delays and thus by available \ac{ADC} technology.  However, this limitation is not relevant for applications requiring a control bandwidth of $\le 1$~MHz, and we find that in such an application, the \ac{FPGA} servo performs similarly to an analog servo.  Because of its ability to realize complex transfer functions that can be controlled and reconfigured remotely, the \ac{FPGA} servo presents significant advantages to an analog controller.

\begin{acknowledgments} 

We wish to thank Arthur K.~Mills for helpful discussions and a close reading of this manuscript.  We acknowledge financial support from the Natural Sciences and Engineering Research Council of Canada (NSERC / CRSNG), the Canadian Foundation for Innovation (CFI), and the Intel \ac{FPGA} University Program.  This work was done under the auspices of the Center for Research on Ultra-Cold Systems (CRUCS).  Finally, we note that the commercial products used in this manuscript are identified for technical clarity, but such use and identification does not necessarily imply our endorsement.

\end{acknowledgments}

\bibliographystyle{osajnl}
\bibliography{FPGA-Servo_RSI}

%
%

\end{document}